\documentstyle[12pt,axodraw]{article}

\newcommand{\be}{\begin{equation}}
\newcommand{\bea}{\begin{eqnarray}}
\newcommand{\ba}{\begin{array}}
\newcommand{\bean}{\begin{eqnarray*}}
\newcommand{\ee}{\end{equation}}
\newcommand{\eea}{\end{eqnarray}}
\newcommand{\ea}{\end{array}}
\newcommand{\eean}{\end{eqnarray*}}

\def\C{L}
\def \dsl {\partial \kern-.55em{/}}
\def \Dsl {D \kern-.65em{/}}
\def \qsl {q \kern-.45em{/}}
\def \slp {p \kern-.45em{/}}
\def \psl {p \kern-.45em{/}}

\def\theequation{\arabic{section}.\arabic{equation}}

\begin{document}

\begin{flushright}
CERN-TH/99-25\\
\end{flushright}

\begin{center}
{\Large {\bf Asymptotic properties of Born-improved amplitudes}}\\[0.4cm]
{\Large {\bf with gauge bosons in the final state}}\\[2.4cm]
{\large Joannis Papavassiliou} \\[0.4cm]
{\em Theory Division, CERN, CH-1211 Geneva 23, Switzerland}\\[0.3cm]

\end{center}

\vskip0.7cm     \centerline{\bf   ABSTRACT}  \noindent

For processes with gauge bosons in the final state
we show how to continuously connect with a single 
Born-improved amplitude
the resonant region, where resummation effects are important,
with the asymptotic region far away from the resonance, where
the amplitude must reduce to its tree-level form.
While doing so
all known field-theoretical constraints are respected, most notably
gauge-invariance, unitarity and the equivalence theorem.
The calculations presented are based on 
the process $f\bar{f}\to ZZ$,
mediated by a possibly resonant Higgs boson; 
this process captures all the
essential features, and can serve as a prototype for a variety of
similar calculations. By virtue of massive
cancellations the resulting 
closed expressions for the differential and total
cross-sections are particularly compact. 

\vskip1.0cm
PACS numbers: 11.15.Lk,~11.10.Jj,~14.70.Hp,~14.80.Bn
\newpage

\section{Introduction}
\indent

The physics of unstable particles   in general \cite{THUN}
 and the computation  of
resonant transition amplitudes in particular 
\cite{PHUN}
has attracted significant
attention  in  recent  years, because  it   is both phenomenologically
relevant   and  theoretically  challenging.    The   main  theoretical
difficulty arises from   the fact that  in the  context of non-Abelian
gauge theories the standard Breit-Wigner  resummation used for regulating
physical amplitudes near resonances is  at odds with gauge invariance,
unitarity, and  the equivalence theorem \cite{EqTh}.
Consequently,  the resulting
Born-improved  amplitudes in general   fail to capture faithfully  the
underlying dynamics.

A solution to this problem has been accomplished at the one-loop level
\cite{PP2,PPHiggs}
by resorting to the reorganisation  of perturbation theory implemented
by the pinch technique (PT) \cite{PT,PT2}.  
The resummation formalism based on the latter
method satisfies a set of crucial  physical requirements, and provides
a  self-consistent framework   for  dealing  with resonant  transition
amplitudes.   The main  thrust of   this diagrammatic  method is  to
exploit  the properties built  into  physical amplitudes in
order   to construct off-shell   Green's  functions with the following
properties 
(i) they are independent of  the gauge-fixing  parameter; 
(ii) they satisfy naive (ghost-free) tree-level  Ward identities (WI's) 
instead of the usual Slavnov-Taylor  identities;     
(iii) they display    physical
thresholds  only \cite{PP2}; 
(iv)  they  satisfy  individually the optical   
and
equivalence theorems \cite{PP2,PRW,PPHiggs}; 
(v) they are analytic functions of the kinematic
variables; 
(vi)  the   effective two-point functions  constructed  are
universal (process-independent) \cite{NJW0},  
Dyson-resummable \cite{PP2,NJW2},  and do not  shift
the  position of the  gauge-independent complex  pole \cite{PP2,KPAS2}.

From   the phenomenological point  of  view  the  upshot  of the above
framework
is  to  construct  Born-improved  amplitudes in  which   all
relevant physical information has been  encoded.  
This in turn is 
useful for the detailed study of the physical properties of particles,
most  importantly the correct  extraction of their masses, widths, and
line shapes.  
The precise measurement  of the mass and  the width of the
$W$ gauge boson for example is of fundamental physical importance.  At
present $W$  bosons are produced at the  Fermilab Tevatron (single $W$
production) and at the CERN LEP2  ($W$ pair-production), whereas large
numbers of $W$ bosons are expected to  be produced at the Large Hadron
Collider (LHC).  In addition,  muon colliders are scheduled to operate
as Higgs factories  for  intermediate energies  of about  500 GeV, and
copious amounts of  Higgs bosons through resonant s-channel production
are expected \cite{FMC}.

Given the   importance  of Born-improved amplitudes,  one   must study
their properties further. One 
open  question in this context
is  how to connect  smoothly  resonant with   asymptotic regions.  
 On
physical grounds   one  expects 
that     far from the   resonance   the
Born-improved   amplitude must behave exactly  as  its  tree-level
counterpart; in fact, 
a  self-consistent resummation formalism should  have
this property built in, i.e. far from resonance one should 
recover the correct high energy behaviour without having to
to  re-expand  the Born-improved amplitude   perturbatively. 
Recovering the correct asymptotic behaviour is particularly tricky
however when the final particles are gauge bosons.
In order
to accomplish  this,  
in addition   to   the correct  one-loop
(running) width, the
appropriate  one-loop vertex corrections  must   be supplemented;
these vertex corrections and the width must be related 
by a crucial tree-level Ward identity.
In practice this WI
ensures   that
massive  cancellations which  take place at  tree-level  
will still go  through after the Born-amplitude has been
``dressed''.

The  need for preserving  tree-level   Ward identities 
when dealing with gauge bosons in the final state
has been emphasized from various points of view in the 
recent  literature.  
As was  pointed   out first  in \cite{BZ} in the 
context of the process   $q\bar{q}' \to \ell \nu  \gamma$, maintaining
the  electromagnetic gauge invariance   associated  with the  outgoing
photon necessitates such a  WI  relating the running  (fermionic) width
coming   from the $W$   self-energy and  the  $WW\gamma$  vertex
containing a fermionic triangle. 
The phenomenological implications of this observation 
were further studied in
\cite{nine}, where the complete
set of fermionic corrections was taken into account
\footnote{In the fermionic case 
both 
the gauge-fixing parameter independence of the result
and
the preservation of the WI
are automatic, because these corrections are Abelian-like.}. 
The non-Abelian case has been addressed in  \cite{PP2} using the PT;
the non-trivial point in this context
is   to  construct  one-loop running  widths  and   
one-loop three-boson vertices which are
independent  of  the  gauge-fixing parameter,  and  at  the  same time
satisfy tree-level  WI.  \footnote{In the   context of the  Background
Field Method \cite{BFM} or the axial gauges \cite{axial}
only the latter property is guaranteed.} Finally, as was
shown in   \cite{PPHiggs} these  WI are   crucial  for satisfying  the
equivalence theorem before and after resummation.  However, to date it
has not been demonstrated explicitly ({\bf i}) 
how the  need for maintaining the
tree-level WI manifests itself at the level  of the cross-section, for
both    Abelian (fermionic)  {\it and}  non-Abelian  (bosonic)  corrections,
({\bf ii}) what is the precise field-theoretical mechanism which restores 
the correct high-energy behaviour; as we will see in detail,
the WI by itself is neither a necessary nor a sufficient 
condition for recovering the correct asymptotic behaviour,
and must be combined with additional requirements, 
({\bf iii}) whether the PT algorithm 
has all aforementioned necessary requirements built in it.
 
In     this paper we   will address the issues listed above.
In particular, we will 
show with
detailed analytic calculations
that   the Born-improved amplitudes
constructed by means of the PT resummation  algorithm not only encodes
correctly the effects on and around the resonant but also far from it.
To study the above points we will  calculate  the  
(resonant) Higgs boson
contributions  to the process $f \bar{f}  \to Z Z$. The motivation for
turning to this particular process is three-fold. First,
from  the theoretical  point  of  view this  process  contains all
necessary features,  without additional technical  complications;  for
example, unlike the  $f \bar{f} \to W^{+}  W^{-}$ it does  not contain
any non-resonant  (non-Higgs boson  related) background  due to an 
$s$-channel $\gamma$ and $Z$.   Second, the Higgs boson self-energy
and vertex  receives  contributions  from loops   containing fermions,
scalars,   and gauge bosons.  Therefore  this  process  can  serve as a
prototype for studying the relevant issues. Third,
the   resonant process $f  \bar{f} \to   Z Z$    may be
intrinsically  interesting for muon  colliders,  if the Standard Model
Higgs boson  turns out to be heavier  than $2M_Z$. 
Therefore, the exact closed expressions for the Born-improved amplitude
presented here may be useful
for studying various properties of this process in detail.   

The  paper is organized as  follows:  In section II we derive  closed
expressions for  the  differential and total cross-sections  away from
the resonance, and establish their high energy behaviour. In section III
we calculate the same cross-sections where we now account for resonant
effects; in  particular we derive closed  expressions for an arbitrary
running width and and an  arbitrary form of
the $HZZ$ vertex compatible with Lorentz
invariance and CP symmetry.  We then analyse in detail 
the mechanism which enforces the correct high energy behaviour of the
Born-improved amplitude when the PT width and vertex are used.
In section IV 
we
show how  the mechanism
of  the  previous section can  be  realized
explicitly in the  PT   context.  
In  section V we   present    our
conclusions.  Finally, various  useful  formulae are  presented in an
Appendix.
Throughout  this  work  we only   consider  the absorptive (imaginary)
contributions to the width and the vertex; they constitute the
minimum amount of ``dressing''
necessary in  order to  regulate the resonant
amplitude.  The real (dispersive)  corrections can also be included in
a systematic way, but  this task is  beyond the  scope of  the present
work.

\setcounter{equation}{0}
\section{The differential and total cross-sections}
In this section we present closed expressions
for the tree-level 
differential and total cross-sections for the process 
$f(p_1)\bar{f}(p_2)\to  Z(k_1)Z(k_2)$ and study their behaviour
in the limit where the center-of-mass energy is much larger 
than any other mass scale. The purpose is twofold: (i) we show
that if the cancellations of the PT are implemented before
the calculation the resulting expressions are rather compact
(ii) Based on these closed expressions we can easily establish
the behaviour of the cross-section far from resonance.

The
tree-level  transition   amplitude ${\cal T}_{\mu\nu}$  for the   process
$f(p_1)\bar{f}(p_2)\to  Z(k_1)Z(k_2)$ is the  sum of an $s$-, a
$t$-, and a $u$- channel contribution (Fig.\ 1), denoted   
by ${\cal T}_{s\, \mu\nu}$, ${\cal T}_{t\, \mu\nu}$,
and ${\cal T}_{u\, \mu\nu}$, 
respectively, given by
\begin{eqnarray}
  \label{THsZZ}
{\cal T}_{s\, \mu\nu} \ &=&\ \bar{v}(p_2) \Gamma^{Hf\bar{f}}_0 u(p_1)
\Delta_{0} (s)\,\Gamma^{HZZ}_{0\mu\nu}
\, ,\nonumber \\
  \label{TtZZ}
{\cal T}_{t\, \mu\nu} \ &=&\ \bar{v}(p_2) 
\Gamma^{Zf\bar{f}}_{0\nu}\, \frac{1}{\not\! p_1 - \not\! k_1 - m_f}\, 
\Gamma^{Zf\bar{f}}_{0\mu}\, u(p_1) \, ,\nonumber\\ 
 \label{TuZZ}
{\cal T}_{u\, \mu\nu} \ &=&\ \bar{v}(p_2) 
\Gamma^{Zf\bar{f}}_{0\mu}\, \frac{1}{\not\! p_1 - \not\! k_2 - m_f}\,
                            \Gamma^{Zf\bar{f}}_{0\nu}\, u(p_1)\, .
\end{eqnarray}

Here,   $s=(p_1+p_2)^{2}=(k_1+k_2)^2$  is the   center-of-mass 
energy squared,
$\Gamma^{HZZ}_{0\mu\nu}   = (ig_w\,     M_Z/c_w ) g_{\mu\nu}$, 
$\Gamma^{Hf\bar{f}}_0   =    -i    g_w\,  m_f    /   (2    M_W)$   and
$\Gamma^{Zf\bar{f}}_{0\mu}\,  =\, -ig_w/(2c_w)\, \gamma_\mu\, [  T^f_z
(1 - \gamma_5) - 2Q_fs^2_w]$, with $c_w = \sqrt{1 - s^2_w} = M_W/M_Z$,
are  the   tree-level  $HZZ$,  $Hf\bar{f}$  and $Zf\bar{f}$ couplings,
respectively, $Q_f$ is the electric charge of the fermion $f$, and
$T^f_z$ its $z$-component of the weak isospin.
Away from the resonance the propagator
for the Higgs boson is given by the usual tree-level
expression $\Delta_{0} (s) = (s-M_H^2)^{-1}$.
For on shell $Z$ bosons, i.e. $k_{1}^{2}=k_{2}^{2}=M_Z^2$,
the vertex $\Gamma^{HZZ}_{0\mu\nu} $ satisfies the following
WI
\begin{eqnarray}
k_1^{\mu}k_2^{\nu}\Gamma^{HZZ}_{0\mu\nu}=
\frac{ig_w\,M_Z}{2c_w}[\Delta_{0}^{-1}(s)+ (M_H^2-2M_Z^2) ],  
\label{BWI}
\end{eqnarray}
which, as we will see, controls the high energy behaviour of the tree-level 
amplitude.

The Mandelstam variables $t$ and $u$ are given by
\bea
t &=(k_1-p_1)^2=(p_2-k_2)^2=
-\frac{1}{4}(\beta_Z^2+\beta^2_f-2zx)s~, \nonumber\\
u &= (k_1-p_2)^2=(p_1-k_2)^2= 
-\frac{1}{4}(\beta_Z^2+\beta^2_f+2zx )s~, 
\eea
where $x \equiv \cos\theta$ is the center-of-mass scattering angle,
and
\be
\beta_Z = \sqrt{1-\frac{4M^2_Z}{s}} ~, ~~~~~~~
\beta_f = \sqrt{1-\frac{4m^2_f}{s}} ~, ~~~~~~~ 
z=\beta_Z\beta_f~.
\ee

The squared matrix element 
$\overline{|{\cal M}|^2}$
averaged over initial and final
polarization states is given by
\be
\overline{|{\cal M}|^2} = 
\frac{1}{4}\sum_{s_1,s_2}
 \left[ \bar{\upsilon } T_{\mu\nu} u \right]
\left( g^{\mu\mu '} - \frac{k^{\mu}_1 k^{\mu '}_1}{M_Z^2}\right)
\left( g^{\nu\nu '} - \frac{k^{\nu}_2 k^{\nu '}_2}{M_Z^2}\right)
\left[ \bar{u} T_{\mu '\nu '}^{\dagger} \upsilon \right]~, \\
\label{M2}
\ee
and
the unpolarized differential cross section for leptons in the
initial state 
\footnote{If the initial fermions are quarks we must multiply by 
a factor of $\frac{1}{3}$.}
reads
\be
\frac{d\sigma}{dx} = \frac{1}{32\pi}\frac{\beta_Z}{\beta_f} 
\frac{1}{s}
\overline{|{\cal M}|^2}~. 
\ee

For the actual calculation it is convenient to
write
$\overline{|{\cal M}|^2}$ as the sum of six
sub-amplitudes distinguished by their dependence on the
three Mandelstam variables $s$, $t$, and $u$.
In carrying out this decomposition we follow
the method explained in detail in \cite{PRW};
in particular, we carry out analytically a large number
of cancellations between terms originating from the
longitudinal pieces $k^{\mu}_{1} k^{\mu '}_{1} / {M_Z^2}$
and $k^{\nu}_{2} k^{\nu '}_{2}/{M_Z^2}$ appearing on the 
left-hand side of Eq.(\ref{M2}).
These cancellations are carried out systematically 
by resorting to the PT reorganisation of the amplitude, i.e. 
we use the 
tree-level Ward identity obeyed by the amplitude
in order
to extract 
$s$-channel-like pieces from $t$ and $u$ graphs, which
cancel against analogous contributions coming from 
the usual $s$-channel graph.
Specifically, we start from the following
elementary WI satisfied by the two sub-amplitudes: 
\begin{eqnarray}
k_1^{\mu}k_2^{\nu}{\cal T}_{s\, \mu\nu}
&=& 
{\cal T}_{s} +{\cal T}_{P}~, \nonumber\\
k_1^{\mu}k_2^{\nu}({\cal T}_{t\, \mu\nu}+{\cal T}_{u\, \mu\nu})
&=& 
M_Z^2({\cal T}_{t} + {\cal T}_{u}) - {\cal T}_{P}~,
\label{EWI}
\end{eqnarray}
with
\be
{\cal T}_{P} = \bar{v}(p_2)
\Gamma^{Hf\bar{f}}_0 u(p_1)\Delta_{0}(s)
\Bigg(\frac{ig_w M_Z}{2c_w}\Bigg) \Delta_{0}^{-1}(s)~,
\label{TP}
\ee
and
\begin{eqnarray}
{\cal T}_{s}&=&
\ \bar{v}(p_2) \Gamma^{Hf\bar{f}}_0 u(p_1)
\Delta_{0} (s)\, \Bigg(\frac{ig_w M_Z}{2c_w}\Bigg) 
(M_H^2-2M_Z^2)~, \nonumber\\
{\cal T}_{t} &=& 
\bar{v}(p_2) 
\Gamma^{G^{0}f\bar{f}}_{0}\, \frac{1}{\not\! p_1 - \not\! k_1 - m_f}\, 
\Gamma^{G^{0}f\bar{f}}_{0}\, u(p_1) \, ,
\nonumber\\
{\cal T}_{u} &=& 
\bar{v}(p_2) 
\Gamma^{G^{0}f\bar{f}}_{0}\, \frac{1}{\not\! p_1 - \not\! k_2 - m_f}\, 
\Gamma^{G^{0}f\bar{f}}_{0}\, u(p_1) \, ,
\end{eqnarray}
where
$\Gamma^{G^{0}f\bar{f}}=-\, g_w\,(m_f/M_W)\, T^f_z \gamma_5$ 
is the coupling of the neutral Goldstone boson $G^{0}$ 
to the fermions. Then, by
adding both parts of Eq.(\ref{EWI}) 
we see that the  
${\cal T}_{P}$ terms cancel on the RHS and we are left exactly with 
what one expects from
the (generalized) equivalence theorem \cite{PPHiggs}.

After carrying out the above  cancellations, 
a straightforward calculation shows that the   
differential
cross-section reads
\be
\frac{d\sigma}{dx} =
\Bigg( \frac{\pi}{64}\Bigg)
\Bigg( \frac{\alpha_w^2}{c_w^4}\Bigg)
\Bigg (\frac{\beta_Z}{\beta_f}\Bigg )\Bigg (\frac{1}{s} \Bigg) 
\Bigg[ M_{ss}+M_{st}+M_{su}+M_{tt}+M_{uu}+M_{ut} \Bigg]~,
\ee
where $\alpha_w=g^2_w/(4\pi)$, and
\bea
M_{ss} &=& 
s\Delta_0^2 
m_f^2\beta^2_f f_1 ~,\nonumber\\ 
M_{st} &=&  
\frac{2s}{(t-m_f^2)}\Delta_0 m_f^2
\left[f_2\beta_f^2 - f_3zx\right]~,\nonumber\\ 
M_{su} &=&
\frac{2s}{(u-m_f^2)}\Delta_0 m_f^2
\left[ f_2\beta_f^2 +f_3zx\right] ~,\nonumber \\ 
M_{tt} &=&
\frac{1}{2} 
\frac{s^2}{(t-m_f^2)^2}
\left[ f_4z^2 x^2 - 
f_5zx+ f_6   \right]~, \nonumber\\
M_{uu} &=&
\frac{1}{2}
\frac{s^2}{(u-m_f^2)^2}
\left[ f_4z^2 x^2 +
f_5zx + f_6 \right]~,\nonumber\\
M_{ut} &=&
\frac{1}{2}
\frac{s^2}{(u-m_f^2)(t-m_f^2)}
\left[ f_7z^2 x^2 +f_8 \right]~,
\label{dcs}
\eea
with
\bea
f_1 &=& 12 - 8\frac{s}{M_Z^2}
+4 \frac{M_H^2}{M_Z^2}+\frac{M_H^4}{M_Z^4}~,\nonumber\\
f_2 &=& 3-4a_f-\frac{s}{M_Z^2}~,\nonumber \\
f_3 &=& 1+4a_f-\frac{s}{M_Z^2}
+\frac{m_f^2}{M_Z^4}\left(M_H^2+2M_Z^2\right)~,\nonumber \\
f_4 &=& -\left(2a_f^2+3a_f+\frac{1}{8}\right)
-\frac{1}{2}\frac{m_f^2}{M_Z^2}\left( 4a_f+1\right)
-\frac{1}{2}\frac{m_f^4}{M_Z^4}~,\nonumber\\
f_5 &=& \frac{m_f^2}{s}\left(8a_f^2-4a_f+ \frac{9}{2}\right) ~,
\nonumber\\
f_6 &=&\left(2a_f^2+3a_f+\frac{1}{8}\right)\beta_Z^2+
\frac{m_f^2}{s}\left(8a_f^2-28a_f+ \frac{5}{2}\right)
+\frac{1}{2}\frac{m_f^2}{M_Z^2}\left(4a_f+1\right)\nonumber\\
&&-\frac{m_f^2M_Z^2}{s^2}\left(48a_f^2-56a_f+3\right) 
-\frac{2m_f^4}{s^2}(16a_f^2-16a_f+3)+\frac{1}{2}\frac{m_f^4}{M_Z^4}
-\frac{2m_f^4}{M_Z^2s}~,\nonumber\\
f_7 &=& \frac{m_f^2}{M_Z^2}\left(1-4a_f\right)+\frac{m_f^4}{M_Z^4}~,
\nonumber\\
f_8 &=& \frac{m_f^2}{s}\left(16a_f^2-56a_f-3\right) 
+\frac{16M_Z^2}{s}\left( 2a_f^2+3a_f+\frac{1}{8}\right)
+\frac{m_f^2}{M_Z^2}\left(4a_f-1\right)\nonumber\\
&&-\frac{4m_f^4}{s^2}\left(16a_f^2-16a_f-1\right)
-\frac{2m_f^2M_Z^2}{s^2}\left(16a_f^2+56a_f+1\right)
-\frac{m_f^4}{M_Z^4}+\frac{4m_f^4}{sM_Z^2}~,\nonumber\\
&& ~
\label{fi}
\eea
where we have defined $a_f= {(T^f_z-2Q_fs_w^2)}^2$.

The total cross-section is given by
\be
\sigma = \frac{1}{2!} \int_{-1}^{1} dx \Bigg(\frac{d\sigma}{dx}\Bigg)~,
\ee
where 
the statistical factor $\frac{1}{2!}$ is due to the
two identical $Z$ bosons in the final state.
After carrying out the angular integration 
$\sigma$ is given by 
\be
\sigma= \Bigg(\frac{\pi}{128}\Bigg) \Bigg(\frac{\alpha_{w}^{2}}{c_w^4}\Bigg)
\Bigg( \frac{\beta_Z}{\beta_f}\Bigg)\Bigg(\frac{1}{s}\Bigg) 
\Bigg[\sigma_{ss}+\sigma_{st}+\sigma_{su}+ \sigma_{tt}
+\sigma_{uu}+\sigma_{ut}\Bigg]~,
\label{tcs}
\ee
where
\bea
\sigma_{ss} &=& 2 s\Delta_0^2 m_f^2\beta^2_f f_1~,
\nonumber\\
\sigma_{st} =\sigma_{su} &=&
-8\Delta_0 m_f^2 
\left[f_3 - \frac{V}{2z}(\beta^2_ff_2 + y f_3)\right]~,
\nonumber\\
\sigma_{tt}= \sigma_{uu} &=&
\frac{4}{y^2-z^2}\left[z^2f_4 +yf_5+f_6 \right] + 8f_4
-\frac{2}{z} (f_5+2yf_4)V~, 
\nonumber\\
\sigma_{ut} &=& \frac{2}{zy}\left[z(z-2y)f_7 + (y^2f_7+f_8)V \right]~,
\eea
and 
\be 
y=-\frac{1}{2}(1+\beta_Z^2)~,~~~~~
V=\ln |\frac{y+z}{y-z}|~,
\ee  
and we have used that $t-m_f^2 = (s/2)(y+zx)$ and
$u-m_f^2 = (s/2)(y-zx)$.

\medskip
\medskip

The following 
comments are now in order:

\begin{itemize}

\item[(\bf i)]
By virtue of the extensive cancellations described at the beginning
of this section,
the resulting expressions for the differential and total cross
sections are particularly compact. 

\item[(\bf ii)]
For $m_f=0$
the differential and total cross sections given above 
reduce to the expressions given   
in Eq.(3.5) and Eq.(3.7) of \cite{BM}, respectively.

\item[(\bf iii)]
Notice that all sub-amplitudes $f_i$ given in Eq.\ (\ref{fi})
behave at most as constants for large $s$. This is a generic
feature of the PT reorganisation of the amplitude,
as was demonstrated first in \cite{PRW} for the case of
$e^{+}e^{-} \to W^{+}W^{-}$.

\item[(\bf iv)]
It is straightforward to verify that
in the limit $s\gg \mu^2$, where $\mu$ is any
of the particle masses in the process, i.e.
$m_f^2$, $M_z^2$, $M_W^2$, and $M_H^2$,
we have that
$f_2=f_3=\frac{1}{8}f_1$, $f_5=0$, $f_4=-f_6$ and  
$f_7=-f_8$. 
In that limit we obtain 
\begin{eqnarray}
\sigma_{ss}+\sigma_{st}+\sigma_{su} &=&   
{\cal  O} (\frac{\mu^2}{s})~, \nonumber\\   
\sigma_{tt}+\sigma_{uu}+\sigma_{ut} &=& 8 |f_4|\ln(s/m_f^2) +\dots~,
\label{limit}
\end{eqnarray}
where the ellipsis denote terms which are at most constants.
Consequently, for large $s$ the total cross section is
the (manifestly positive) quantity
\be
\sigma = \Bigg(\frac{\pi}{16}\Bigg)
\Bigg (\frac{\alpha_{w}^{2}}{c_w^4}\Bigg) 
\Bigg(\frac{1}{s}\Bigg) |f_4| \ln(s/m_f^2)~. 
\label{asbeh}
\ee
We see that, as is expected on physical grounds, 
the asymptotic behaviour of the cross-section is
determined by the ``genuine''
$t$-and $u$-channel terms, i.e. the $t$-and $u$-channel 
terms remaining after
the cancellations of the longitudinal polarization momenta 
has been carried out.

\item[(\bf v)]
The expression for $\sigma_{ss}$ is identical to 
the imaginary (absorptive) part of the gauge-invariant 
set of one-loop self-energy-like graphs involving
two virtual $Z$ bosons 
(together with the corresponding Goldstone bosons and ghosts)
derived in \cite{PPHiggs}, given also in section IV of the present paper.

\end{itemize}

\setcounter{equation}{0}
\section{The Born-improved amplitude}

In  this section we will recompute the amplitude for the
process  $f(p_1)\bar{f}(p_2)\to  Z(k_1)Z(k_2)$   using a      
generic
parametrization  for the width of the Higgs boson,
and for the $HZZ$ vertex. This calculation will  
show quantitatively 
how the high energy behaviour of the amplitude is altered
if the parametrization  of the width and the vertex is kept
arbitrary, and the precise role of the WI will be analysed.
In addition it will be shown that
if the PT width and vertex are used, the correct
high energy behaviour will emerge.

In the vicinity of the Higgs boson resonance,
i.e. for $s\sim M_H^2$, the 
amplitude ${\cal T}_{s\, \mu\nu}$ given in Eq.~(\ref{THsZZ})
diverges, and must be regulated by introducing 
a width in the Higgs boson propagator. In particular
we must replace the tree-level $\Delta_{0}$ by a $\Delta$
of the form
\be
\Delta = [s-M^2_H + i \Im m\ \Pi(s)]^{-1}~,
\label{HPropRes}
\ee
where $\Pi(s)$ is the (appropriately defined)
one-loop self-energy of the Higgs boson.
For the purposes of this work
it is convenient to introduce the dimensionless quantity $\C(s)$
as follows:
\be
\Im m\ \Pi(s) =   s\C(s) .
\label{defL}
\ee
The
correct one-loop expressions for the various decay channels
contributing to $\Im m\ \Pi (s)$ have  
been derived in \cite{PPHiggs} and are also reported
in the next section.
However for the purposes of this calculation $\C$ will be
treated as an arbitrary parameter.
Similarly,
the most general tensorial decomposition of the 
$HZZ$ vertex, where the two on-shell $Z$  are assumed
to be contracted
with their corresponding polarization vectors, reads 
\bea
  \label{HZZdec}
{G}^{HZZ}_{\mu\nu}(q,p,k)\ &=&\ \frac{ig_w M_Z }{c_w}\,  
\Big[\, (1+A(s))\, g_{\mu\nu}\, +\, 
B(s)\, \frac{q_\mu\, q_\nu}{q^2}\, \Big]\nonumber\\
&=& \Gamma^{HZZ}_{0\mu\nu}+ \widehat{\Gamma}^{HZZ}_{\mu\nu}~.
\label{Gparam}
\eea
Notice that in general the CP-violating  form-factor proportional to 
$\epsilon_{\mu\nu\rho\sigma}k_{1}^{\rho} k_{2}^{\sigma}$ 
may appear in Eq.(\ref{Gparam}), but it vanishes 
at one-loop in  the Standard Model
\footnote{There are also contributions originating from the one-loop 
mixing between the Higgs boson and the $Z$ boson.
Such contributions have been treated correctly 
within the PT framework in \cite{KPAS1, APRL}.  
It is easy to verify from  
the expressions given in Eq.6 of the first reference in \cite{APRL} 
that all such mixing contributions
are non-resonant in the entire range of the 
relevant phase-space.}.
The explicit
one-loop expressions (see Fig.\ 2)
for $A$ and $B$ are also computable by means of the PT,
and will be presented in the next section,
but for the purposes of this section they too will 
be treated as arbitrary quantities.

Next we calculate the differential and total cross-sections
using the above modified propagator and vertex.
The new differential cross section is obtained from Eq.\ (\ref{dcs})
after replacing ${M}_{ss}$, ${M}_{st}$ and ${M}_{su}$ by the
the modified amplitudes $\widehat{M}_{ss}$, 
$\widehat{M}_{st}$, and $\widehat{M}_{su}$, respectively, given by:
\bea
\widehat{M}_{ss}  &=&
 s|{\Delta}|^2 m_f^2 \beta^2_f \widehat{f}_1 \nonumber~,\\ 
\widehat{M}_{st}&=& 
\frac{2s^2}{(t-m_f^2)} |{\Delta}|^2 m_f^2  
\left[ \widehat{f}_2\beta_f^2 - \widehat{f}_3zx \right ]~,
\nonumber\\
\widehat{M}_{su}
&=&
\frac{2s^2}{(u-m_f^2)}|{\Delta}|^2 m_f^2
\left[\widehat{f}_2\beta_f^2 +\widehat{f}_3zx\right]~, 
\eea
where 
\be
|{\Delta}|^2 = [(s-M_H^2)^2+s^2\C^2]^{-1}~,
\ee
and
\bea
\widehat{f}_1 &=& f_1
+\frac{s^2}{M_Z^4}(\C-R)^2 -
4\frac{s}{M_Z^2}[R(A+B)+A\C] +4(3A^2+B^2+2AB)~,
\nonumber\\
\widehat{f}_2 &=& f_2 (1- \frac{M^2_H}{s})
+ B\C + (3-4a_f)A\C -\frac{s}{M_Z^2}\C R~,
\nonumber \\
\widehat{f}_3 &=&
f_3 (1- \frac{M^2_H}{s})
+ (1+4a_f)\C R + 2 \frac{m^2_f}{M^2_Z}A\C
-\frac{s}{M_Z^2}\C\bigg(R - \frac{m^2_f}{M^2_Z}(R-\C)\bigg)~,
\label{widef}
\eea
with 
\be
R\equiv A+\frac{1}{2}B~.
\label{defR}
\ee
Similarly, 
the total cross-section is given from Eq.\ (\ref{tcs})
after replacing $\sigma_{ss}$, $\sigma_{st}$ and $\sigma_{su}$
by $\widehat\sigma_{ss}$, $\widehat\sigma_{st}$, and 
$\widehat\sigma_{su}$, respectively, given by
\bea
\widehat\sigma_{ss} &=& 2 s|{\Delta}|^2
m_f^2\beta^2_f \widehat{f}_1~,
\nonumber\\
\widehat\sigma_{st} =\widehat\sigma_{su} &=&
-8s|{\Delta}|^2 m_f^2
\left[\widehat{f}_3 - \frac{V}{2z}
(\beta^2_f \widehat{f}_2 + y \widehat{f}_3)\right]~.
\eea
Clearly, only the purely $s$-channel contributions together with the
interference terms are modified, while
the ``genuine'' $t$- and $u$- channel contributions 
(box-like terms) remain unaffected.

It is now obvious from Eq.\ (\ref{widef}) that one cannot recover 
the correct high energy behaviour of the amplitude
for generic values of the functions $\C$, $A$, and $B$.
For example, even if we choose 
the $\C$, $A$, and $B$
such that asymptotically $\C = R$,
if the individual $\C$, $A$, and $B$ grow sufficiently fast with $s$
the resulting amplitude has the wrong large-$s$ limit.
Reversing the situation, {\it in general}
even if the individual $\C$, $A$, and $B$ are assumed not to grow
faster than constants, unless we also have that $\C \to R$,   
the resulting total cross-section will behave 
asymptotically at least as a constant, instead of the correct behaviour given
in Eq.~(\ref{asbeh})\footnote{A physically relevant counter-example is the case where 
$A=B=0$ and $\C=const/s$, which will be studied at the end of this section.}. 

Let us now turn to the $\C$, $A$ and $B$
derived within the PT.
\footnote{In what follows we will use ``hats'' to indicate
all such quantities.}
Using the PT \cite{PPHiggs} 
one can reorganize the one-loop $S$-matrix in such a way as to define 
Higgs boson and Goldstone boson self-energies, $\widehat{\Pi}(q^2)$
and $\widehat{\Pi}^{G^0G^0} (q^2)$, respectively,
and $HZZ$ and $HG^{0}G^{0}$ vertices,
$\widehat{\Gamma}_{\mu\nu}^{HZZ}$ and $\widehat{\Gamma}^{HG^0G^0}$,
respectively,
endowed with all the important properties listed in the Introduction.
In particular,
(i)
asymptotically $\widehat{\C}$ goes to a constant, whereas $\widehat{A}$ 
and $\widehat{B}$ 
grow logarithmically (this has been established in \cite{PPHiggs}
and is also studied in detail in the next section),
(ii)
they are related by the
following tree-level WI
\footnote{
To see that Eq.\ (\ref{PTHZZ1}) has 
indeed
the same form as its tree-level counterpart 
notice that 
for $\widehat{\Gamma}_{\mu\nu}^{HZZ} \to \widehat{\Gamma}_{0\mu\nu}^{HZZ}$,
$\widehat{\Pi}(q^2) \to (q^2-M_H^2)$, 
$\widehat{\Gamma}^{HG^0G^0}\to \Gamma^{H G^{0}G^{0}}_{0}= 
ig_w M_Z M_H^2/{2c_w}$, and $\widehat{\Pi}^{G^0G^0}(M_Z^2)\to M_Z^2$
one recovers the WI of Eq.\ (\ref{BWI}).}  
\be
\label{PTHZZ1}
k^\mu_1 k^\nu_2 \widehat{\Gamma}_{\mu\nu}^{HZZ}\, 
+\, M^2_Z \widehat{\Gamma}^{HG^0G^0}
= \frac{ig_wM_Z}{2c_w}\, \Big[\, \widehat{\Pi}(q^2)\, +\, 
\widehat{\Pi}^{G^0G^0}(k^2_1)  
+\, \widehat{\Pi}^{G^0G^0}(k^2_2)\, \Big]\, ~.
\ee

We are now in position to study 
explicitly 
how the
correct high energy behaviour of the Born-improved amplitude is enforced.
First of all, by virtue of the first property listed above, 
i.e. because the individual 
$\widehat{\C}$, $\widehat{A}$, and $\widehat{B}$ grow mildly with
$s$, we only need to show that $\C = R $ asymptotically.
To see how this comes about,
we start with the parametrization of
$\widehat{\Gamma}_{\mu\nu}^{HZZ}$ given in Eq.\ (\ref{HZZdec})
and act with $k^\mu_1 k^\nu_2$ on both sides;
in the limit $s\gg M^2_Z$, we obtain 
\be
k^\mu_1 k^\nu_2 \widehat{\Gamma}_{\mu\nu}^{HZZ}
= \ \frac{ig_wM_Z}{2c_w}~s \widehat{R}~.
\label{WIver1}
\ee
On the other hand, putting the $Z$'s on shell in Eq.\ (\ref{PTHZZ1})
and using the form of  $\widehat{\Pi}^{HH}(s)$
given in Eq.\ (\ref{HPropRes}), we have that 
\be
\label{WIver2}
k^\mu_1 k^\nu_2 \widehat{\Gamma}_{\mu\nu}^{HZZ}
=\frac{ig_wM_Z}{2c_w}\, 
\Big[\, \widehat{\C} s + 2 \widehat{\Pi}^{G^0G^0}(M^2_Z) \Big]  
- M^2_Z \widehat{\Gamma}^{HG^0G^0}~.
\ee
Setting 
\be
\widehat{\Gamma}^{HG^0G^0}= 
\frac{ig_wM_Z}{2c_w}\, \widehat{D}(s)~,
\label{D}
\ee
we obtain after
equating the left-hand sides of Eq.\ (\ref{WIver1}) and 
Eq.\ (\ref{WIver2})
\be
\widehat{R}-\widehat{\C} = 
\frac{1}{s}\Big[2 \widehat{\Pi}^{G^0G^0}(M^2_Z)-M^2_Z \widehat{D} (s)\Big]~.
\label{RC}
\ee
Thus, the role of the WI is to supply this last relation;
however,
no additional information about the high energy behaviour
of either side of Eq.\ (\ref{RC}) is provided.

The next important step is to establish that
the terms inside the parenthesis on the left-hand side 
of Eq.\ (\ref{RC})
grow at most logarithmically
for large values of $s$. Indeed, to begin with, 
$\widehat{\Pi}^{G^0G^0}(M^2_Z)$ is 
a constant, independent of $s$.
On the other hand, $\widehat{D}(s)$ has a non-trivial dependence on $s$; 
using the diagrams of Fig.\ 3, together with 
the Feynman rules given in \cite{DDW}
\footnote{Here we are using the well-known fact 
\cite{BFMPT}
that at one-loop 
the PT effective Green's functions
coincide with the conventional (gauge-fixing-parameter- dependent)
Green's functions 
of the Background Field Method, for the special value 
$\xi_Q=1$ . This correspondence does not persist
beyond one loop \cite{NJW2loops}.} and Eq.~(A.6)
in the Appendix, 
one can verify that 
\be
\widehat{D}(s) \sim  \ln(s/\mu^2) + \dots~,
\ee
and thus, from  Eq.\ (\ref{RC})
\be
\widehat{R}-\widehat{\C} = 0 + {\cal O}\bigg(s^{-1}\ln(s/\mu^2)\bigg)~.
\ee

As mentioned before,
this last relation is crucial for 
 recovering the correct asymptotic behaviour for
the amplitude.
Indeed, in the limit $s\gg \mu^2$, setting $\C=R$
in Eq.\ (\ref{RC}) yields
\bea
\widehat{f}_1 &=& - 8\frac{s}{M_Z^2}(1+\widehat{R}^2) + ...~,
\nonumber\\
\widehat{f}_2 &=& -\frac{s}{M_Z^2}(1+\widehat{R}^2) + ...~,
\nonumber \\
\widehat{f}_3 &=& -\frac{s}{M_Z^2}(1+ \widehat{R}^2) + ...~.
\eea
Thus, 
as happens in the non-resonant case of the previous section
$\widehat{f}_2 =  \widehat{f}_3 = \frac{1}{8}\widehat{f}_1$
and therefore
\be
\widehat\sigma_{ss}+\widehat\sigma_{st}+\widehat\sigma_{su}=
{\cal O}\bigg( s^{-1}\ln(s/\mu^2)\bigg)~.
\ee

We are now in a position to fully appreciate the role
of the WI. 
Even though 
asymptotically $\widehat{\C}$ goes to a constant, whereas $\widehat{A}$ 
and $\widehat{B}$ 
grow logarithmically, 
(a fact which, without additional information,  
would make us infer that the high energy behaviour of the amplitude 
would be
distorted)
delicate cancellations
make the crucial quantity $(\C - R)$ energetically suppressed.
Thus, cancellations taking place on the
left-hand side 
of Eq.\ (\ref{RC}), whose study would necessitate explicit 
knowledge of $\widehat{\C}$, $\widehat{A}$ and $\widehat{B}$,
are directly encoded in the $\widehat{D}$ appearing on the 
right-hand side.

The conclusion of this analysis is that 
by virtue of the one-loop Ward identities
{\sl and} the good individual high-energy behaviour of the
PT one-loop Green's functions
the tree-level and Born-improved amplitudes
coincide sufficiently far
away from the resonance region.

We end this section by studying  
the behaviour of the 
Born-improved
amplitude
for two different choices
for the parameters $A$, $B$ and $\C$; these choices are not just 
arbitrary mathematical 
examples, but have instead a rather well-known 
field-theoretical origin.
In the first case the WI is violated but the correct 
high energy behaviour of the amplitude
is nonetheless recovered; in the second case
the WI is fullfiled, but 
the high energy behaviour is distorted.
The first example corresponds to the case where  
a constant width is used in the Born-improved amplitude, i.e.
the
Higgs boson width has the form 
$\Im m\ \Pi = \sum_i c_{i}\theta(s-4M^2_i) $, 
 where the coefficients $c_i$ are constants, independent of $s$;
their exact expressions may be obtained from
Eq. (\ref{widths}) by setting $s=M_H^2$. 
In addition we make the simplest assumption for
the vertex, namely $A=B=0$; clearly, such an assumption
violates the WI given in Eq.\ (\ref{BWI}).
For large values of $s$ we than have that  
$\Im m\ \Pi  = \sum_i c_{i}\equiv c$.
In that case we have that $\C= c/s$ and the 
corresponding expressions for the $\widehat{f}_1$, 
$\widehat{f}_2$, and $\widehat{f}_3$ reduce to 
\be
\widehat{f}_1 = f_1 +\frac{c^2}{M_Z^4} ,~~~
\widehat{f}_2 = f_2 (1- \frac{M^2_H}{s}),~~~
\widehat{f}_3 =
f_3 (1- \frac{M^2_H}{s})+\frac{c^2}{M^4_Z} \frac{m^2_f}{s}~.
\label{widefconwid}
\ee
So, in the limit of large $s$ we recover the correct asymptotic relation
for the $\widehat{f}_i$, and therefore for the entire cross-section.
Of course, the use of a constant width 
is known to be problematic for other reasons, for example the
fact that it leads to a violation of the optical theorem \cite{OT}
(for details see the second paper of \cite{PP2}).

The second case pertains to the unitary
gauge (the $\xi\to \infty$ limit of the renormalizable $R_{\xi}$ gauges). 
In this gauge the WI of Eq. (\ref{PTHZZ1}) is satisfied
by the conventional two-and three point functions, before
resorting to the PT algorithm \cite{WI2}; however, their 
imaginary parts display a strong dependence on $s$.
For example 
the running width corresponding to two virtual $W$
bosons in the unitary gauge
is given by $\Im m\ {\Pi}^{(\infty)} 
\sim (g^2/M_W^2)(s^2-4sM_W^2+12 M_W^4$),
and so, for large $s$ we have that $\C \sim (g^2/M_W^2)s$, which 
leads to a gross distortion of
the high energy behaviour of the Born-improved amplitude.

\setcounter{equation}{0}
\section{Explicit realization in the pinch technique framework.} 

In this section we will show explicitly that the PT self-energies
and vertices satisfy the required relations at high energies.
In particular we will prove the validity of Eq.\ (\ref{RC})
without resorting to the WI of  Eq.\ (\ref{PTHZZ1})
as we did in the previous section, 
but instead by 
showing directly that, for asymptotic values of $s$, 
$\widehat{\C} = \widehat{R}$. 
This calculation 
constitutes
a non-trivial test of the entire construction; for practical 
purposes
it is essential,
given the fact that
the Green's functions
related to the unphysical Goldstone bosons,
while they are crucial for realising the WI of Eq.\ (\ref{PTHZZ1}),
do not explicitly appear
in the actual
computation of 
the cross-section. Indeed, the only quantities 
which appear in 
Eq.\ (\ref{widef}) 
are the $\widehat{A}$, $\widehat{B}$, and $\widehat{\C}$, 
 but not the $\widehat{D}$ of Eq. (\ref{D}) nor the $\widehat{\Pi}^{G^0G^0}$.

We next proceed with the explicit calculation.
The partial running widths for the Higgs boson 
have been first 
calculated at one-loop in \cite{PPHiggs}; they are given by:
\begin{eqnarray}
\Im m\ \widehat{\Pi}_{(WW)}(s) & =& 
\frac{\alpha_w}{16}\frac{M_H^4}{M_W^2}
\Big[\, 1+4\frac{M_W^2}{M_H^2}- 4\frac{M_W^2}{M_H^4}
(2s-3M_W^2)\, \Big]\beta_W \theta (s-4M_W^2)\, , \nonumber\\ 
\Im m\ {\widehat \Pi}_{(ZZ)}(s) &=& \frac{\alpha_w}{32}
\frac{M^4_H }{M^2_W}\Big[\, 1+4\frac{M^2_Z }{M^2_H}
-4\frac{M^2_Z }{M^4_H}(2s-3M^2_Z)\Big]\, \beta_Z \theta (s-4M_Z^2)\, ,
\nonumber\\
\Im m\ {\widehat \Pi}_{(FF)}(s) &=& N_{F}
\frac{\alpha_w}{8}
\frac{m_{F}^2}{M^2_W}s\beta_{F}^3 
\theta (s-4m_{F}^2)\, ,\nonumber\\
\Im m\ {\widehat \Pi}_{(HH)}(s) &=& \frac{9\alpha_w}{32}\frac{M^4_H }{M^2_W}
\beta_H \theta (s-4M_H^2)\, .
\label{widths}
\end{eqnarray}
In the above formula we denote by $F$ the various
fermionic flavours appearing inside the quantum loops,
i.e. $F \in \{ e,\mu ,\tau , u, d, c, s, t, b \}$.
$N_{F}=1$ for leptons, and $N_{F} =3$ for quarks.
In the case of a heavy Higgs boson the channels which dominate
numerically are the $WW$, $ZZ$ and $tt$ .
From the above expressions we can extract the dimensionless quantities
$\widehat{\C}^{(WW)}$, $\widehat{\C}^{(ZZ)}$, $\widehat{\C}^{(FF)}$, and
$\widehat{\C}^{(HH)}$,
according to the definition of Eq.\ (\ref{defL}); in the limit of large $s$ they
will be simply the coefficients multiplying $s$ in the expressions 
given on the right-hand side of Eq.\ (\ref{widths}) .

The absorptive form-factors 
$\widehat{A}$  and $\widehat{B}$  of the $HZZ$ one-loop vertex 
are obtained from the graphs shown in Fig.\ 2, using the Feynman rules of 
\cite{DDW}, and can be expressed
in terms of the standard Passarino-Veltman 
one-loop integrals \cite{tHV} as given in \cite{BAK}. 
For on shell external $Z$ bosons 
the arguments of the $C$ functions appearing in the calculation are
$C(M_Z^2,M_Z^2,s,m_i^2,m_j^2,m_i^2)$, where $m_i$, $m_j$ 
are the masses of the particles inside the triangle,
$(iji) \in \{ (FFF), (WWW), (ZHZ), (HZH) \}$.
We will use the shorthand 
notation where the first three (common) arguments will be suppressed,
and the remaining three masses will be denoted as a superscript, i.e.
$C(M_Z^2,M_Z^2,s,m_i^2,m_j^2,m_i^2) \equiv C^{(iji)}$.
Similarly, for the $B_0$ functions we use the shorthand notation
$B_0 (s,m_i,m_i)\equiv B_0^{(ii)}$. Finally, a ``bar'' over 
$B_0$ and $C$ indicates that only their absorptive 
part has been considered. 
The individual diagrams yield:
\footnote{
The closed expression for the PT absorptive form-factors 
$\widehat{A}$  and $\widehat{B}$
have been presented first in \cite{PPHiggs}, but
here we correct several misprints.}

\begin{eqnarray}
  \label{AZa}
i \widehat{A}_{(a)} &=& N_{F}
\frac{\alpha_w}{8\pi}\, \frac{m^2_t}{M^2_W}\, \Big\{ a_{F}\, 
\Big[ 
s\beta_Z^2(\bar{C}_0^{(FFF)}+2\bar{C}_{11}^{(FFF)})\, 
-\, 8\bar{C}_{24}^{(FFF)}\Big]
\nonumber\\
&& +\, \frac{1}{4}\Big[ s(\beta_Z^2+2\beta_{F}^2)
\bar{C}_0^{(FF)}\, +\, 
2s\beta^2_Z\bar{C}_{11}^{(FFF)}\, 
-\, 8\bar{C}_{24}^{(FFF)} \Big]\Big\},\nonumber\\
  \label{AZb1}
i \widehat{A}_{(b1)} &=& -\, \frac{2\alpha_w}{\pi}\, \frac{M^4_W}{M^4_Z}\, 
\bar{B}_0^{(WW)}\, ,\nonumber\\
  \label{AZb2c5c6}
i\widehat{A}_{(b2)} &=&  -\, \frac{\alpha_w}{16\pi}
\frac{3}{2}\frac{M^2_H}{M^2_W}\bar{B}_0^{(HH)}\, ,\nonumber\\
  \label{AZb3}
i\widehat{A}_{(b3)} &=& -\, \frac{\alpha_w}{16\pi} \Big[\Big(
\frac{M^2_H}{M^2_W}+2\Big)\Big( 2\frac{M^2_W}{M^2_Z} -1\Big)^2
\bar{B}_0^{(WW)}
+\, \frac{1}{2}\Big(\frac{M^2_H}{M^2_W} +\, 2\frac{M^2_Z}{M^2_W}\Big) 
\bar{B}_0^{(ZZ)}\,\Big]\, ,\nonumber\\
  \label{AZb4}
i\widehat{A}_{(b4)} &=& \frac{\alpha_w}{\pi}\, \frac{M^4_W}{M^4_Z}\,
\bar{B}_0^{(WW)}\, ,\nonumber\\
  \label{AZc1}
i\hat{A}_{(c1)} &=& \frac{2\alpha_w}{\pi}\, \frac{M^4_W}{M^4_Z}\,
\Big[ 4\bar{C}_{24}^{(WWW)}+ (s-2M_Z^2)\bar{C}_0^{(WWW)}\Big]\, ,\nonumber\\
  \label{AZc2c3}
i\widehat{A}_{(c2)}+i\widehat{A}_{(c3)} &=& -\frac{\alpha_w}{\pi}\,
\frac{M^2_W}{M^2_Z}\, s\, \bar{C}_0^{(WWW)}\, ,\nonumber\\
  \label{AZc4}
i\widehat{A}_{(c4)} &=& \frac{\alpha_w}{4\pi}\, \Big[2M^2_W
\bar{C}_0^{(WWW)}\, +\, \frac{M^4_Z}{M^2_W}
\bar{C}_0^{(ZHZ)}\Big]\, ,\nonumber\\
\label{AZc7}
i\widehat{A}_{(c5)} &=& i\widehat{A}_{(c6)}\ =\ 0\, ,\nonumber\\
i\widehat{A}_{(c7)} &=&  \frac{\alpha_w}{8\pi}\, \Big[ 2(M^2_H+2M^2_W)
\bar{C}_0^{(WWW)} +\, 3M^2_Z\frac{M^2_H}{M^2_W}\bar{C}_0^{(HZH)}\Big]\, ,
\nonumber\\
  \label{AZc8}
i\widehat{A}_{(c8)} &=&   \frac{\alpha_w}{4\pi}\, \Big[ 
\Big(2\frac{M^2_W}{M^2_Z}-1\Big)^2\Big(\frac{M^2_H}{M^2_W}+2\Big)
\bar{C}_{24}^{(WWW)}\, \nonumber\\
&&+\  \frac{1}{2}\,
\Big(\frac{M^2_H}{M^2_W} +\, 2\frac{M^2_Z}{M^2_W}\Big)
\bar{C}_{24}^{(ZHZ)}\, +\, \frac{3}{2}\, \frac{M^2_H}{M^2_W}
\bar{C}_{24}^{(HZH)}\Big]\, ,\nonumber\\
  \label{AZc9c10} 
i\widehat{A}_{(c9)}+i\widehat{A}_{(c10)}
&=& -\, \frac{4\alpha_w}{\pi}\, \frac{M^4_W}{M^4_Z}\, 
\bar{C}_{24}^{(WWW)}\, .
\label{theA}
\end{eqnarray}
 
The individual contributions  to the $B$  form-factor are
given by 
\begin{eqnarray}
  \label{BZa}
i\widehat{B}_{(a)} &=& -\, N_{F}
\frac{\alpha_w}{4\pi}\, \frac{m^2_{F}}{M^2_W}s\, 
\Big\{ a_{F}\, [3\bar{C}_0^{(FFF)}  
+ 4\bar{C}_{11}^{(FFF)}
-4\bar{C}_{23}^{(FFF)}] \\
&&+\, \frac{1}{4}[\bar{C}_{0}^{(FFF)}-
4\bar{C}_{23}^{(FFF)}]\Big\} ,\nonumber\\
  \label{BZb1234c47}
i\widehat{B}_{(b1)} &=& i\widehat{B}_{(b2)}\ 
=\ i\widehat{B}_{(b3)}\ =\ i\widehat{B}_{(b4)}\
=\ i\widehat{B}_{(c4)}\ =\ i\widehat{B}_{(c7)}\ =\ 0\, ,\nonumber\\
  \label{BZc1}
i\widehat{B}_{(c1)} &=& -\, \frac{4\alpha_w}{\pi }s\frac{M^4_W}{M^4_Z}\, 
[2\bar{C}_{12}^{(WWW)}+2\bar{C}_{23}^{(WWW)}+\bar{C}_0^{(WWW)}]
\, ,\nonumber\\
  \label{BZc2c3}
i\widehat{B}_{(c2)}+i\widehat{B}_{(c3)} &=& -\frac{2\alpha_w}{\pi}\, 
s\frac{M^2_W}{M^2_Z}\,\bar{C}_{11}^{(WWW)}\, ,\nonumber\\
  \label{BZc5c6}
i\widehat{B}_{(c5)}+i\widehat{B}_{(c6)} &=& -\, 
\frac{\alpha_w}{2\pi}\frac{s}{M^2_Z}\, \Big[
2(2M^2_W-M^2_Z)\bar{C}_{12}^{(WWW)}
 -\, \frac{M^4_Z}{M^2_W}\, \bar{C}_{12}^{(ZHZ)}\Big],  
\nonumber\\
  \label{BZc8}
i\widehat{B}_{(c8)} &=& -\, \frac{\alpha_w}{4\pi}\,s \Big[\, 
\Big(2\frac{M^2_W}{M^2_Z}-1\Big)^2 \Big(\frac{M^2_H}{M^2_W}+2\Big)
(\bar{C}_{12}^{(WWW)}+\bar{C}_{23}^{(WWW)})\nonumber\\
&&+\, \frac{1}{2}\Big(\frac{M^2_H}{M^2_W}+2\frac{M^2_Z}{M^2_W}\Big)
(\bar{C}_{12}^{(ZHZ)}+\bar{C}_{23}^{(ZHZ)})
 +\, \frac{3}{2}\, \frac{M^2_H}{M^2_W}\, (\bar{C}_{12}^{(HZH)}+
\bar{C}_{23}^{(HZH)})\, \Big]\, ,\nonumber\\
\label{BZc9c10}
i\widehat{B}_{(c9)} + i\widehat{B}_{(c10)} &=& \frac{4\alpha_w}{\pi}\,
s\frac{M^4_W}{M^4_Z}\, [\bar{C}_{12}^{(WWW)}+\bar{C}_{23}^{(WWW)}]\, . 
\label{theB}
\end{eqnarray}
 
In deriving the above results
we have also used the identity 
\be
\bar{C}_0^{(iji)} + \bar{C}_{11}^{(iji)} + \bar{C}_{12}^{(iji)} =0 ~.
\ee
.

From the Eq.\ (\ref{theA}) and Eq.\ (\ref{theB}) we may collect
the total contribution of each separate channel to $\widehat{A}$ and 
$\widehat{B}$, which will be denoted
by $\widehat{A}^{(iji)}$ and 
$\widehat{B}^{(iji)}$; using the definition in Eq.\ (\ref{defR})
we may then construct the 
corresponding $\widehat{R}^{(iji)}$. 
In order to determine their asymptotic behaviour,
we must use that
in the limit of large $s$
\be
\bar{C}_{11}^{(iji)} \to  -\frac{1}{s} \bar{B}_{0}^{(ii)},~~   
\bar{C}_{23}^{(iji)} \to   \frac{1}{s}\bar{B}_{0}^{(ii)},~~ 
\bar{C}_{24}^{(iji)} \to  \frac{1}{4s} \bar{B}_{0}^{(ii)},
\ee
as one may easily verify 
using the formulae
presented in the Appendix.
Thus we arrive at the following limits for the various $\widehat{R}$:
\bea
&& \widehat{R}^{(FFF)} 
\to  N_{F}\frac{\alpha_w}{8\pi}\frac{m^2_{F}}{M^2_W}
\bar{B}_0^{(FF)} \to 
\frac{\alpha_w}{8}\frac{m^2_{F}}{M^2_W} = \widehat{\C}^{(FF)}~,\nonumber\\
&& \widehat{R}^{(WWW)} \to -\frac{\alpha_w}{2\pi}\bar{B}_0^{(WW)} 
\to -\frac{\alpha_w}{2} = \widehat{\C}^{(WW)}~,\nonumber \\
&& \widehat{R}^{(ZHZ)} 
\to -\frac{\alpha_w}{4\pi}\frac{M^2_Z}{M^2_W}\bar{B}_0^{(ZZ)} 
\to -\frac{\alpha_w}{4}\frac{M^2_Z}{M^2_W} = \widehat{\C}^{(ZZ)}~,\nonumber\\
&& \widehat{R}^{(HZH)} \to 0 = \widehat{\C}^{(HH)}~.
\eea
This is the announced result.
We notice that all necessary cancellations which lead to
the desired result take place  
channel by channel, as one would 
expect on physical grounds. 
It is also important to emphasize that 
the explicit expressions for the $\widehat{\C}$,
$\widehat{A}$ and $\widehat{B}$ derived in this section 
allow for a detailed study 
of the amplitude 
for the entire range of $s$, and not only
asymptotically, as we have done here.

\setcounter{equation}{0}
\section{Conclusions} 

In this paper we have we shown that within  the PT resummation formalism
the resonant and asymptotic regions  of processes with gauge bosons in
the final state can be described correctly and connected to each other
smoothly by means of a single Born-improved amplitude.  In particular,
using the resonant process $f\bar{f}\to ZZ$ as a reference process, we
have  studied  in  detail  the  mechanism  which enforces the correct
high-energy  behaviour of  the   Born-improved amplitude, and we  have
shown how this mechanism is  in fact automatically (but non-trivially)
realized in the PT framework. This    provides   an      additional
self-consistency check for the resummation formalism  based on the PT.

An  important  by-product  of this  analysis   is that explicit closed
expression   for  the  differential  and  total  cross-sections of the
process  $f\bar{f}\to ZZ$ have been computed,  both at the tree-level 
and
in  the Born-improved approximation. For  the  latter case the generic
form of the amplitude for arbitrary self-energy and vertex-corrections
has  been reported, as well as  the specific corrections obtained from
the PT effective Green's function.
Of course,  if $M_H > 2 M_Z$
the channel $f \bar{f} \to
W^{+}  W^{-}$   will  be  also relevant;  however,  the
 analysis presented in this paper may  be     carried out
straightforwardly to the latter process, with the additional technical
complications of computing the $Z$ and $\gamma$ mediated background.

In the present   work we have only  treated  the case  where the (two)
gauge bosons appeared in the final state. The above considerations may
be generalised to the case where  both incoming and outgoing particles
are gauge bosons.  
The $W$-fusion sub-process $WW  \to ZZ$ for example has
been recently  studied   for the case of off-shell $W$'s 
\cite{KPWJS}; it
was shown how the  tree-level PT
rearrangement of the process $qq\to qqZZ$  
restores the  good  high energy behaviour of the aforementioned
sub-process, thus solving a 
long-standing problem \cite{RKWJS}. 
Based   on the analysis  presented
here  one expects  
that    the   results   
established 
in \cite{KPWJS}
 will  persists  after the one-loop PT
corrections necessary for regulating the $W$-fusion amplitude 
near the Higgs boson resonance have been included.

Given the explicit results  for the process $f\bar{f}\to ZZ$ presented
in  this  paper one   could carry  out a    detailed study of   the 
Standard Model
Higgs boson line shape, obtained from the above  process.  Such a study
could be of potential interest in  the context of a muon-collider, for
example, but is beyond the scope of the present paper.

\newpage

\vspace{0.7cm}\noindent {\bf  Acknowledgments.}  This work has been
funded by  a Marie Curie  Fellowship (TMR-ERBFMBICT 972024).   
I thank R.~Pittau 
for independently checking several of
the calculations appearing in this paper, 
and A.~Pilaftsis and K.~Philippides for various useful discussions.

\def\theequation{\Alph{section}.\arabic{equation}}
\setcounter{equation}{0}
\begin{appendix}
 
\section{Absorptive parts of the $B_0$ and $C$ functions}

In this Appendix we list some formulae which are useful when computing the
absorptive (imaginary) parts of the $B_0$ and $C$ functions.

\medskip
 
For the imaginary part of the $B_0$ function we have (Fig.\ 4a)

\be
\bar{B}_{0}(q^2,m_1^2,m^2_2)= 
\frac{1}{q^2}
\pi\, \theta [ q^2-(m_1+m_2)^2 ]\,\lambda^{1/2}(q^2,m^2_1,m^2_2)~,
\ee
where $\lambda (x,y,z) = (x-y-z)^2 - 4yz$.
In the cases studied in this paper we have 
always $m_1=m_2=m$, and the above formula reduces to
\be
\bar{B}_{0}(q^2,m^2,m^2)= \pi\, \theta (q^2-4m^2 ) \beta ~.
\ee

\medskip

The imaginary parts of $C_0$ in the general case are given by (Fig.\ 4b)

\be
\bar{C}_0= \frac{\pi}{2}\, \theta\bigg(q^2_1-(m_2+m_3)^2\bigg) 
\lambda^{-\frac{1}{2}}(q^2_1,M_2^2,M_3^2)
\ln\Bigg[\frac{\rho^{-}_{1}-m_1^2}{\rho^{+}_{1}-m_1^2}\Bigg] + {\rm c.p.}~,
\ee
with
\be
\rho^{\pm}_1= M_2^2+m_3^2-
\frac{1}{2q_1^2}\bigg[(q_1^2+M_2^2-M_3^2)(q_1^2+m_3^2-m_2^2)
\pm \lambda^{\frac{1}{2}}(q^2_1,M_2^2,M_3^2)
\lambda^{\frac{1}{2}}(q^2_1,m_2^2,m_3^2)\bigg]~,
\ee
and the abbreviation c.p. means cyclic permutation with respect to
1,~2,~3.

\medskip

For the particular channels appearing in our calculations we have:

\bea
\bar{C}_0^{(WWW)}&=& 
\frac{\pi}{2}\,\frac{1}{s\beta_Z} 
\theta (s-4M_W^2)
\ln \bigg(\frac{1+\beta^2_Z-2\beta_Z\beta_W}
{1+\beta^2_Z+2\beta_Z\beta_W}\bigg)~,
\nonumber\\
\bar{C}_0^{(FFF)} &=& 
\frac{\pi}{2}\,\frac{1}{s\beta_Z} 
\theta (s-4m_{F}^2)
\ln\bigg (\frac{1+\beta^2_Z-2\beta_Z\beta_{F}}
{1+\beta^2_Z+2\beta_Z\beta_{F}}\bigg)~,
\nonumber\\
\bar{C}_0^{(HZH)}&=& 
\frac{\pi}{2}\,\frac{1}{s\beta_Z} 
\theta (s-4M_H^2)
\ln\bigg(\frac{1+\beta^2_H-2\beta_Z\beta_H}
{1+\beta^2_H+2\beta_Z\beta_H}\bigg)~,
\nonumber\\
\bar{C}_0^{(ZHZ)} &=& 
\frac{\pi}{2}\,\frac{1}{s\beta_Z} 
\theta (s-4M_Z^2)\ln\bigg(\frac{1-\beta_H^2}
{1+4\beta_Z^2-\beta_H^2}\bigg)~.
\eea

For large $s$ the above formulae reduce to
\be
\bar{C}_0^{(iji)}=
-\frac{\pi}{2}\,\theta (s-4M_i^2)\frac{1}{s} \ln(s/M_i^2)~.
\ee
\newpage

Finally, using the formulae of \cite{BAK},
the exact expressions for the remaining $C$ functions 
are given by

\bea
\bar{C}_{12}^{(FFF)} &=& 
\frac{1}{s\beta_Z^2}
\bigg( M_Z^2 \bar{C}_{0}^{(FFF)} -
\bar{B}_{0}^{ (FF)} \bigg)~,\nonumber\\
\bar{C}_{23}^{(FFF)} &=&  
\frac{1}{s^3\beta_Z^4}
\bigg(
[2M_Z^4(M_Z^2-4m_{F}^2)+2sM_Z^2(M_Z^2+3m_{F}^2)
- s^2(M_Z^2 + m_{F}^2)]
\bar{C}_0^{(FFF)}~,\nonumber\\
&& + (2M_Z^4 - 3sM_Z^2 + s^2)\bar{B}_{0}^{(FF)} \bigg)~,\nonumber\\
\bar{C}_{24}^{(FFF)}&=&
\frac{1}{s\beta_Z^2}
\bigg( [2m_{F}^2 M_Z^2-\frac{1}{2}M_Z^4-\frac{1}{2}s m_{F}^2]
\bar{C}_0^{(FFF)} 
+ [\frac{1}{4}s -\frac{1}{2}M_Z^2 ]\bar{B}_{0}^{(FF)}\bigg)~.
\eea

\bea 
\bar{C}_{12}^{(WWW)} &=& 
\frac{1}{s\beta_Z^2}
\bigg( M_Z^2 \bar{C}_{0}^{(WWW)} - \bar{B}_{0}^{(WW)} \bigg)~,\nonumber\\
\bar{C}_{23}^{(WWW)} &=& 
\frac{1}{s^3\beta_Z^4}
\bigg( [ 2M_Z^4 (M_Z^2 - 4M_W^2) + 2s M_Z^2 (M_Z^2+3 M_W^2)
-s^2(M_Z^2+ M_W^2)] \bar{C}_0^{(WWW)}~,\nonumber\\
&&  + [2M_Z^4 - 3sM_Z^2+ s^2]\bar{B}_{0}^{(WW)} \bigg)~,\nonumber\\
\bar{C}_{24}^{(WWW)} &=&\frac{1}{s\beta_Z^2} \bigg(
[2 M_Z^2 M_W^2 - \frac{1}{2}M_Z^4 -\frac{1}{2}sM_W^2]\bar{C}_{0}^{(WWW)}
+ [\frac{1}{4}s - \frac{1}{2} M_Z^2 ]\bar{B}_{0}^{(WW)} \bigg)~.
\eea

\bea              
         \bar{C}_{12}^{(ZHZ)} &=&\frac{1}{s\beta_Z^2} \bigg(
                M_H^2  C_{0}^{(ZHZ)} - \bar{B}_0^{(ZZ)} \bigg)~,
          \nonumber\\
         \bar{C}_{23}^{(ZHZ)} &=& \frac{1}{s^3\beta_Z^4}
         \bigg(
          [ 2M_Z^2 M_H^2 (M_H^2-4M_Z^2)
           +2sM_H^2 (5M_Z^2-M_H^2)- 2s^2 M_H^2] \bar{C}_{0}^{(ZHZ)} \nonumber\\
&&    + [2M_Z^2 M_H^2 +sM_H^2 - 4sM_Z^2 + s^2] \bar{B}_0^{(ZZ)} \bigg)~, 
\nonumber\\
     \bar{C}_{24}^{(ZHZ)} &=& \frac{1}{s\beta_Z^2}
     \bigg(M_H^2 [2M_Z^2  -\frac{1}{2} M_H^2 - \frac{1}{2}s ] 
     \bar{C}_{0}^{(ZHZ)} 
     + [\frac{1}{2} M_H^2 - M_Z^2 +\frac{1}{4}s]\bar{B}_0^{(ZZ)}\bigg)~.
       \nonumber\\
\eea 

\bea 
       \bar{C}_{12}^{(HZH)} &=& \frac{1}{s\beta_Z^2}
       \bigg ( (2M_Z^2 - M_H^2)\bar{C}_{0}^{(HZH)} - \bar{B}_0^{(HH)}\bigg)~,
        \nonumber\\ 
       \bar{C}_{23}^{(HZH)} &=& \frac{1}{s^3\beta_Z^4}
      \bigg( [ 2M_Z^2 M_H^2(M_H^2-4M_Z^2)
       +2s(2M_Z^2 M_H^2+3M_Z^4-M_H^4) \nonumber\\
       && + s^2 (M_H^2-3M_Z^2)] \bar{C}_0^{(HZH)}
       + [4M_Z^4 - 2 M_Z^2 M_H^2 -2sM_Z^2 -sM_H^2+ s^2] 
       \bar{B}_{0}^{(HH)}\bigg)~,
        \nonumber\\
       \bar{C}_{24}^{(HZH)} &=& \frac{1}{s\beta_Z^2}
        \bigg(
        [2M_Z^2 M_H^2 -  \frac{1}{2} M_H^4 -  
         \frac{1}{2} s M_Z^2] \bar{C}_0^{(HZH)}
         + [ \frac{1}{4}s - \frac{1}{2} M_H^2 ]\bar{B}_{0}^{(HH)}\bigg)~.
\eea          

\end{appendix}

\newpage

\centerline{\bf FIGURES}
 
\begin{center}
\begin{picture}(400,200)(0,0)
\SetWidth{0.8}
 
\ArrowLine(0,130)(30,100)\ArrowLine(30,100)(0,70)
\DashArrowLine(30,100)(50,100){3}\GCirc(65,100){15}{0.8}
\DashArrowLine(80,100)(100,100){3}\GCirc(115,100){15}{0.8}
\Photon(120,115)(140,130){3}{2}\Photon(120,85)(140,70){3}{2}
\Text(10,140)[r]{$f (p_1)$}\Text(10,60)[r]{$\bar{f}(p_2)$}
\Text(160,140)[r]{$Z_\mu (k_1)$}\Text(160,60)[r]{$Z_\nu (k_2)$}
\Text(40,110)[]{$H$}\Text(90,110)[]{$H$}
\Text(65,125)[]{$\widehat{\Delta}(q)$}
\Text(135,100)[l]{$\Gamma^{HZZ}_{0\mu\nu} + \widehat{\Gamma}^{HZZ}_{\mu\nu}$}
\LongArrow(122,124)(130,132)\LongArrow(122,76)(130,68)
 
\Text(65,20)[]{\bf (a)}
 
\ArrowLine(240,120)(270,120)\ArrowLine(270,120)(270,80)
\ArrowLine(270,80)(240,80)\Photon(270,120)(300,120){2}{4}
\Photon(270,80)(300,80){2}{4}
\Text(240,130)[l]{$f$}\Text(240,70)[l]{$\bar{f}$}
\Text(275,100)[l]{$f (p_1-k_1)$}
\Text(300,130)[r]{$Z_\mu$}\Text(300,70)[r]{$Z_\nu$}
 
\Text(270,20)[]{\bf (b)}
 
\ArrowLine(340,120)(370,120)\ArrowLine(370,120)(370,80)
\ArrowLine(370,80)(340,80)\Photon(370,120)(400,120){2}{4}
\Photon(370,80)(400,80){2}{4}
\Text(340,130)[l]{$f$}\Text(340,70)[l]{$\bar{f}$}
\Text(375,100)[l]{$f (p_1-k_2)$}
\Text(400,130)[r]{$Z_\nu$}\Text(400,70)[r]{$Z_\mu$}
 
\Text(370,20)[]{\bf (c)}
 
\end{picture}\\
{\small {\bf Fig.\ 1:} 
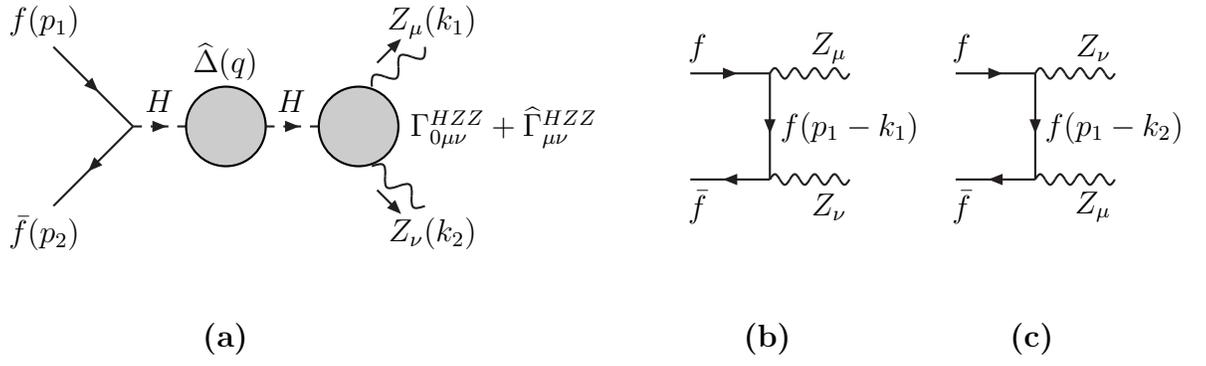
The Born-improved amplitude for the process  
$f\bar{f} \to ZZ$.}
\end{center}
  
\newpage

\begin{center}
\begin{picture}(400,500)(0,0)
\SetWidth{0.8}
 
\DashArrowLine(0,470)(20,470){4}\ArrowLine(20,470)(50,490)
\ArrowLine(50,490)(50,450)\ArrowLine(50,450)(20,470)
\Photon(50,490)(70,490){3}{2.5}\Photon(50,450)(70,450){3}{2.5}
\Text(5,480)[r]{{\small $\widehat{H}(q)$}}
\Text(100,500)[r]{\small $\widehat{Z}_\mu(k_1)$}
\Text(100,440)[r]{\small $\widehat{Z}_\nu(k_2)$}
\Text(30,490)[r]{\small $f,\bar{f}$}
\Text(30,450)[r]{\small $f,\bar{f}$}
\Text(55,470)[l]{\small $f,\bar{f}$}
\Text(35,420)[]{\bf (a)}
 
\DashArrowLine(150,470)(170,470){4}
\PhotonArc(185,470)(15,0,180){3}{5}\PhotonArc(185,470)(15,180,360){3}{5}
\Photon(200,470)(220,490){-3}{2.5}\Photon(200,470)(220,450){3}{2.5}
\Text(185,495)[]{\small $W^+$}
\Text(185,445)[]{\small $W^-$}
\Text(185,420)[]{\bf (b1)}
 
\DashArrowLine(300,470)(320,470){4}
\DashArrowArc(335,470)(15,0,180){3}
\DashArrowArc(335,470)(15,180,360){3}
\Photon(350,470)(370,490){-3}{2.5}\Photon(350,470)(370,450){3}{2.5}
\Text(335,495)[]{\small $H$}
\Text(335,445)[]{\small $H$}
\Text(335,420)[]{\bf (b2)}
 
\DashArrowLine(0,370)(20,370){4}
\DashArrowArcn(35,370)(15,180,360){3}\DashArrowArc(35,370)(15,180,360){3}
\Photon(50,370)(70,390){-3}{2.5}\Photon(50,370)(70,350){3}{2.5}
\Text(35,395)[]{\small $G^+,G^0$}
\Text(35,345)[]{\small $G^-,G^0$}
\Text(35,320)[]{\bf (b3)}
 
\DashArrowLine(150,370)(170,370){4}
\DashArrowArc(185,370)(15,0,180){1.5}\DashArrowArc(185,370)(15,180,360){1.5}
\Photon(200,370)(220,390){-3}{2.5}\Photon(200,370)(220,350){3}{2.5}
\Text(185,395)[]{\small $c^\pm$}
\Text(185,345)[]{\small $c^\pm$}
\Text(185,320)[]{\bf (b4)}
 
\DashArrowLine(300,370)(320,370){4}\Photon(320,370)(350,390){2}{4}
\Photon(350,390)(350,350){2}{4}\Photon(350,350)(320,370){2}{4}
\Photon(350,390)(370,390){3}{2.5}\Photon(350,350)(370,350){3}{2.5}
\Text(345,395)[r]{\small $W^\pm$}
\Text(345,345)[r]{\small $W^\mp$}
\Text(355,370)[l]{\small $W^\pm$}
\Text(335,320)[]{\bf (c1)}
 
\DashArrowLine(0,270)(20,270){4}\Photon(20,270)(50,290){2}{4}
\Photon(50,290)(50,250){2}{4}\DashArrowLine(20,270)(50,250){4}
\Photon(50,290)(70,290){3}{2.5}\Photon(50,250)(70,250){3}{2.5}
\Text(45,295)[r]{\small $W^\pm$}
\Text(45,245)[r]{\small $G^\mp$}
\Text(55,270)[l]{\small $W^\pm$}
\Text(35,220)[]{\bf (c2)}
 
\DashArrowLine(150,270)(170,270){4}\DashArrowLine(170,270)(200,290){4}
\Photon(200,290)(200,250){2}{4}\Photon(170,270)(200,250){2}{4}
\Photon(200,290)(220,290){3}{2.5}\Photon(200,250)(220,250){3}{2.5}
\Text(195,295)[r]{\small $G^\pm$}
\Text(195,245)[r]{\small $W^\mp$}
\Text(205,270)[l]{\small $W^\pm$}
\Text(205,270)[l]{\small $W^\pm$}
\Text(185,220)[]{\bf (c3)}
 
\DashArrowLine(300,270)(320,270){4}\Photon(320,270)(350,290){2}{4}
\DashArrowLine(350,290)(350,250){4}\Photon(350,250)(320,270){2}{4}
\Photon(350,290)(370,290){3}{2.5}\Photon(350,250)(370,250){3}{2.5}
\Text(345,295)[r]{\small $W^\pm,Z$}
\Text(345,245)[r]{\small $W^\mp,Z$}
\Text(355,270)[l]{\small $G^\pm,H$}
\Text(335,220)[]{\bf (c4)}
 
\DashArrowLine(0,170)(20,170){4}\DashArrowLine(20,170)(50,190){4}
\DashArrowLine(50,190)(50,150){4}\Photon(50,150)(20,170){2}{4}
\Photon(50,190)(70,190){3}{2.5}\Photon(50,150)(70,150){3}{2.5}
\Text(45,195)[r]{\small $G^\pm,G^0$}
\Text(45,145)[r]{\small $W^\mp,Z$}
\Text(55,170)[l]{\small $G^\pm,H$}
\Text(35,120)[]{\bf (c5)}
 
\DashArrowLine(150,170)(170,170){4}\Photon(170,170)(200,190){2}{4}
\DashArrowLine(200,190)(200,150){4}\DashArrowLine(170,170)(200,150){4}
\Photon(200,190)(220,190){3}{2.5}\Photon(200,150)(220,150){3}{2.5}
\Text(195,195)[r]{\small $W^\pm,Z$}
\Text(195,145)[r]{\small $G^\mp,G^0$}
\Text(205,170)[l]{\small $G^\pm,H$}
\Text(185,120)[]{\bf (c6)}
 
\DashArrowLine(300,170)(320,170){4}\DashArrowLine(320,170)(350,190){4}
\Photon(350,190)(350,150){2}{4}\DashArrowLine(320,170)(350,150){4}
\Photon(350,190)(370,190){3}{2.5}\Photon(350,150)(370,150){3}{2.5}
\Text(350,195)[r]{\small $G^\pm,H$}
\Text(350,145)[r]{\small $G^\mp,H$}
\Text(355,170)[l]{\small $W^\pm,Z$}
\Text(335,120)[]{\bf (c7)}
 
\DashArrowLine(0,70)(20,70){4}\DashArrowLine(20,70)(50,90){4}
\DashArrowLine(50,90)(50,50){4}\DashArrowLine(20,70)(50,50){4}
\Photon(50,90)(70,90){3}{2.5}\Photon(50,50)(70,50){3}{2.5}
\Text(50,95)[r]{\small $G^\pm,G^0,H$}
\Text(50,45)[r]{\small $G^\mp,G^0,H$}
\Text(55,70)[l]{\small $G^\pm,H,G^0$}
\Text(35,20)[]{\bf (c8)}
 
\DashArrowLine(150,70)(170,70){4}\DashArrowLine(170,70)(200,90){1.5}
\DashArrowLine(200,90)(200,50){1.5}\DashArrowLine(200,50)(170,70){1.5}
\Photon(200,90)(220,90){3}{2.5}\Photon(200,50)(220,50){3}{2.5}
\Text(195,95)[r]{\small $c^\pm$}
\Text(195,45)[r]{\small $c^\pm$}
\Text(205,70)[l]{\small $c^\pm$}
\Text(185,20)[]{\bf (c9)}
 
\DashArrowLine(300,70)(320,70){4}\DashArrowLine(350,90)(320,70){1.5}
\DashArrowLine(350,50)(350,90){1.5}\DashArrowLine(320,70)(350,50){1.5}
\Photon(350,90)(370,90){3}{2.5}\Photon(350,50)(370,50){3}{2.5}
\Text(345,95)[r]{\small $c^\pm$}
\Text(345,45)[r]{\small $c^\pm$}
\Text(355,70)[l]{\small $c^\pm$}
\Text(335,20)[]{\bf (c10)}
\end{picture}\\
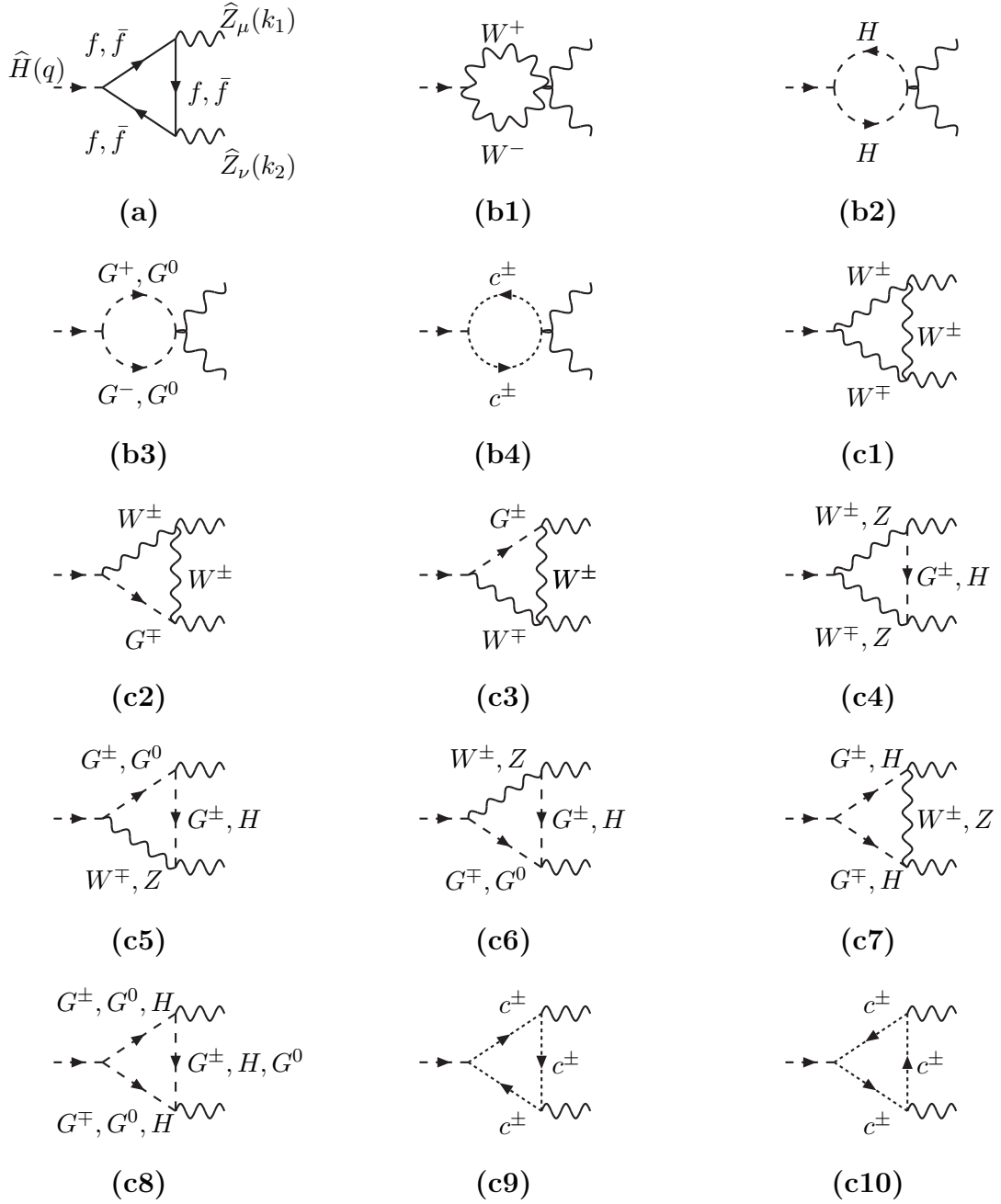
{\small {\bf Fig.\ 2:} ~Diagrams contributing to the one-loop vertex 
$\widehat\Gamma_{\mu\nu}^{HZZ}$.}

\end{center}

\newpage

\begin{center}
\begin{picture}(400,500)(0,0)
\SetWidth{0.8}
 
\DashArrowLine(0,470)(20,470){4}\ArrowLine(20,470)(50,490)
\ArrowLine(50,490)(50,450)\ArrowLine(50,450)(20,470)
\DashArrowLine(50,490)(70,490){4}\DashArrowLine(50,450)(70,450){4}
\Text(5,480)[r]{{\small $\widehat{H}(q)$}}
\Text(100,500)[r]{\small ${\widehat{G}}_0(k_1)$}
\Text(100,440)[r]{\small ${\widehat{G}}_0(k_2)$}
\Text(30,490)[r]{\small $f,\bar{f}$}
\Text(30,450)[r]{\small $f,\bar{f}$}
\Text(55,470)[l]{\small $f,\bar{f}$}
\Text(35,420)[]{\bf (a)}

\DashArrowLine(150,470)(170,470){4}
\PhotonArc(185,470)(15,0,180){3}{5}\PhotonArc(185,470)(15,180,360){3}{5}
\DashArrowLine(200,470)(220,490){4}\DashArrowLine(200,470)(220,450){4}
\Text(185,495)[]{\small $Z,W^+$}
\Text(185,445)[]{\small $Z,W^-$}
\Text(185,420)[]{\bf (b)}
 
\DashArrowLine(300,470)(320,470){4}
\DashArrowArc(335,470)(15,0,180){3}
\DashArrowArc(335,470)(15,180,360){3}
\DashArrowLine(350,470)(370,490){4}\DashArrowLine(350,470)(370,450){4}
\Text(335,495)[]{\small $H$}
\Text(335,445)[]{\small $H$}
\Text(335,420)[]{\bf (c)}
 
\DashArrowLine(0,370)(20,370){4}
\DashArrowArcn(35,370)(15,180,360){3}\DashArrowArc(35,370)(15,180,360){3}
\DashArrowLine(50,370)(70,390){4}\DashArrowLine(50,370)(70,350){4}
\Text(35,395)[]{\small $G^+,G^0$}
\Text(35,345)[]{\small $G^-,G^0$}
\Text(35,320)[]{\bf (d)}
 
\DashArrowLine(150,370)(170,370){4}
\DashArrowArc(185,370)(15,0,180){1.5}\DashArrowArc(185,370)(15,180,360){1.5}
\DashArrowLine(200,370)(220,390){4}\DashArrowLine(200,370)(220,350){4}
\Text(185,395)[]{\small $c_z,c^\pm$}
\Text(185,345)[]{\small $c_z,c^\pm$}
\Text(185,320)[]{\bf (e)}

\DashArrowLine(300,370)(320,370){4}\Photon(320,370)(350,390){2}{4}
\DashArrowLine(350,390)(350,350){4}\Photon(350,350)(320,370){2}{4}
\DashArrowLine(350,390)(370,390){4}\DashArrowLine(350,350)(370,350){4}
\Text(345,395)[r]{\small $Z,W^\pm$}
\Text(345,345)[r]{\small $Z,W^\mp$}
\Text(355,370)[l]{\small $H,G^\pm$}
\Text(335,320)[]{\bf (f)}
 
\DashArrowLine(0,270)(20,270){4}\Photon(20,270)(50,290){2}{4}
\DashArrowLine(50,290)(50,250){4}\DashArrowLine(20,270)(50,250){4}
\DashArrowLine(50,290)(70,290){4}\DashArrowLine(50,250)(70,250){4}
\Text(45,295)[r]{\small $Z$}
\Text(45,245)[r]{\small $G^0$}
\Text(55,270)[l]{\small $H$}
\Text(35,220)[]{\bf (g)}
 
\DashArrowLine(150,270)(170,270){4}\DashArrowLine(170,270)(200,290){4}
\DashArrowLine(200,290)(200,250){4}\Photon(170,270)(200,250){2}{4}
\DashArrowLine(200,290)(220,290){4}\DashArrowLine(200,250)(220,250){4}
\Text(195,295)[r]{\small $G^0$}
\Text(195,245)[r]{\small $Z$}
\Text(205,270)[l]{\small $H$}
\Text(185,220)[]{\bf (h)}
 
\DashArrowLine(300,270)(320,270){4}\DashArrowLine(320,270)(350,290){4}
\Photon(350,290)(350,250){2}{4}\DashArrowLine(350,250)(320,270){4}
\DashArrowLine(350,290)(370,290){4}\DashArrowLine(350,250)(370,250){4}
\Text(345,295)[r]{\small $G^\pm,H$}
\Text(345,245)[r]{\small $G^\mp,H$}
\Text(355,270)[l]{\small $W^\pm,Z$}
\Text(335,220)[]{\bf (i)}
 
\DashArrowLine(0,170)(20,170){4}\DashArrowLine(20,170)(50,190){4}
\DashArrowLine(50,190)(50,150){4}\DashArrowLine(50,150)(20,170){4}
\DashArrowLine(50,190)(70,190){4}\DashArrowLine(50,150)(70,150){4}
\Text(45,195)[r]{\small $G^0,H$}
\Text(45,145)[r]{\small $G^0,H$}
\Text(55,170)[l]{\small $H,G^0$}
\Text(35,120)[]{\bf (j)}
 
\end{picture}\\
{\small {\bf Fig.\ 3:}
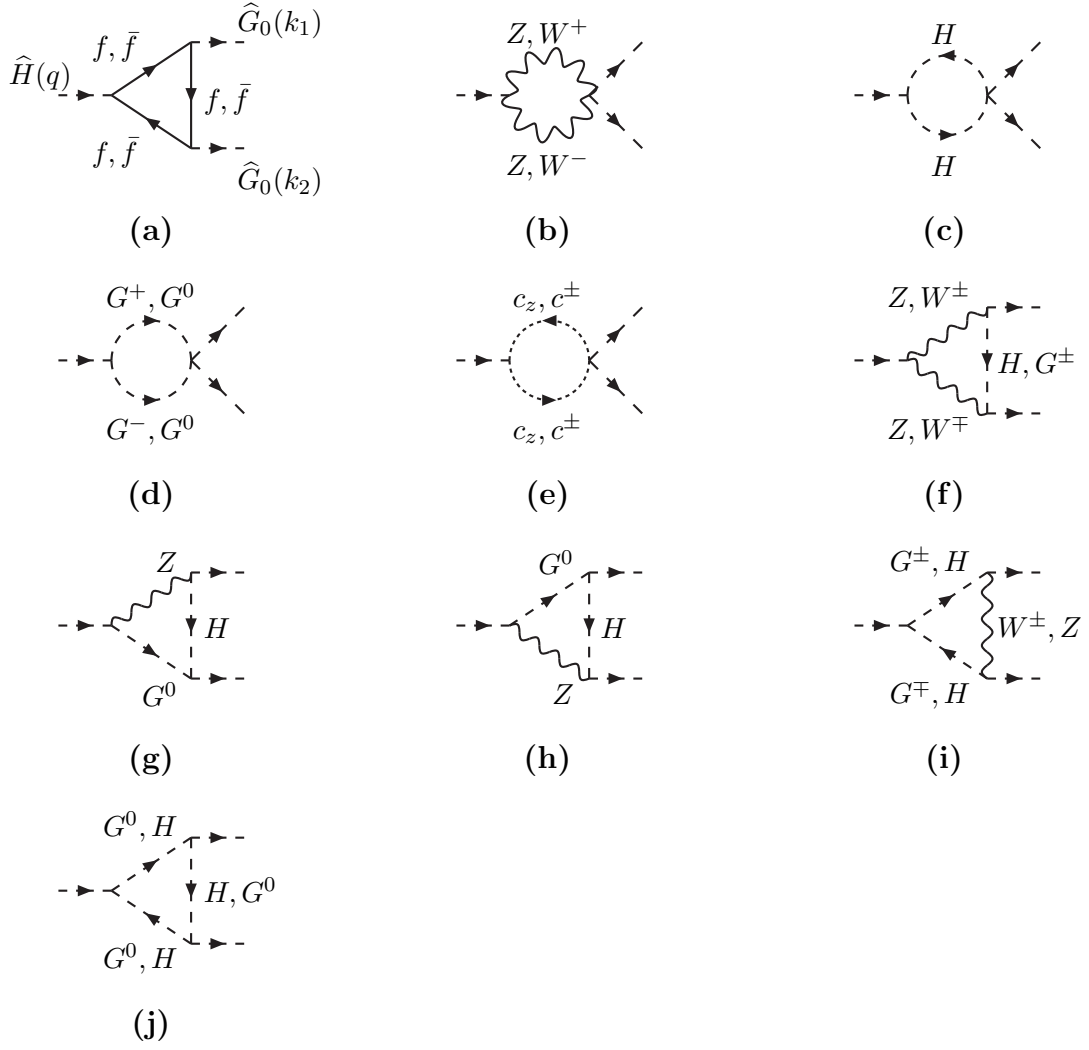
~Diagrams contributing to the one-loop vertex $\widehat\Gamma^{HG^{0}G^{0}}$.}

\end{center}

\newpage

\begin{center}
\begin{picture}(300,100)(0,0)
\SetWidth{0.8}
\Line(90,50)(120,50)  \Text(100,60)[r]{$q$}
\Line(180,50)(210,50) \Text(200,60)[l]{$q$} 
\Text(150,90)[r]{$m_1$}
\Text(150,10)[r]{$m_2$}
\BCirc(150,50){30}
\DashLine(155,90)(155,10){2.} 
\end{picture}\\
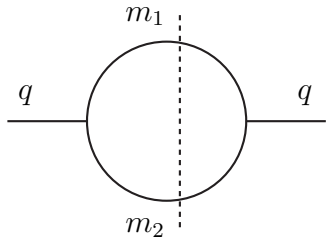
{\small {\bf Fig.\ 4a:} The absorptive part of the two-point function.}
\end{center}

\medskip

\medskip

\medskip

\medskip

\vskip1.7cm 

\begin{center}
\begin{picture}(300,100)(0,0)
\SetWidth{0.8}
\Line(60,50)(100,50)  \Text(90,60)[r]{$q_1,M_1$}
\Line(100,50)(155,90)
\Line(100,50)(155,10)
\Line(155,90)(200,90) \Text(200,100)[r]{$q_2,M_2$}
\Line(155,10)(200,10) \Text(200,20)[r]{$q_3,M_3$}
\Line(155,90)(155,10) \Text(172,52)[r]{$m_1$}
\DashLine(112,70)(112,30){2.}
\DashLine(130,95)(160,60){2.}
\DashLine(130,5)(165,40){2.}
\Text(135,80)[r]{$m_2$}
\Text(130,25)[r]{$m_3$}
\end{picture}\\
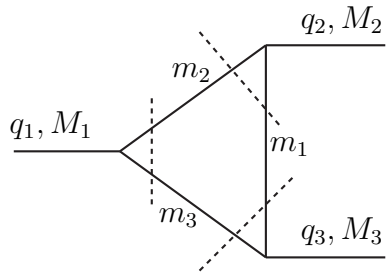
{\small {\bf Fig.\ 4b:} The absorptive parts of the three-point function.}

\end{center}

\end{document}